\documentclass[reprint,superscriptaddress,amssymb,amsmath,aps,prx,longbibliography, nofo otinbib]{revtex4-2}

\usepackage{graphicx}
\usepackage{amssymb}
\usepackage{epstopdf}
\usepackage{upgreek}
\usepackage{dcolumn}
\usepackage{bm}
\usepackage{textcomp, gensymb}  
\usepackage{braket}
\usepackage{amsmath}
\usepackage{mathtools}
\usepackage{float}
\usepackage{tabularx}
\usepackage{mathrsfs}
\usepackage{lineno}
\usepackage{tikz}
\usepackage{graphicx}
\usepackage{sidecap} 
\usepackage{array}
\usepackage{amsthm}
\usepackage{enumitem}
\usepackage{xcolor} 

\expandafter\ifx\csname package@font\endcsname\relax\else
\expandafter\expandafter
\expandafter\usepackage
\expandafter\expandafter
\expandafter{\csname package@font\endcsname}%
\fi

\newcommand{\be}{\begin{eqnarray}}
\newcommand{\ee}{\end{eqnarray}}
\newcommand{\bfig}{\begin{figure}}
\newcommand{\efig}{\end{figure}}

\DeclareFontFamily{U}{mathb}{}
\DeclareFontShape{U}{mathb}{m}{n}{
  <-5.5> mathb5
  <5.5-6.5> mathb6
  <6.5-7.5> mathb7
  <7.5-8.5> mathb8
  <8.5-9.5> mathb9
  <9.5-11.5> mathb10
  <11.5-> mathbb12
}{}

\usepackage[colorlinks=true,allcolors=blue]{hyperref}%

\begin{document}

\title{Bandwidth and visibility improvement in detection of a weak signal using mode entanglement and swapping}

\author{Y. Jiang}
\thanks{These two authors contributed equally}
\affiliation{JILA, National Institute of Standards and Technology and the University of Colorado, Boulder, Colorado 80309, USA}
\affiliation{Department of Physics, University of Colorado, Boulder, Colorado 80309, USA}

\author{E. P. Ruddy}
\thanks{These two authors contributed equally}
\affiliation{JILA, National Institute of Standards and Technology and the University of Colorado, Boulder, Colorado 80309, USA}
\affiliation{Department of Physics, University of Colorado, Boulder, Colorado 80309, USA}

\author{K. O. Quinlan}
\affiliation{JILA, National Institute of Standards and Technology and the University of Colorado, Boulder, Colorado 80309, USA}
\affiliation{Department of Physics, University of Colorado, Boulder, Colorado 80309, USA}

\author{M. Malnou}
\affiliation{National Institute of Standards and Technology, Colorado 80305, USA}

\author{N. E. Frattini}
\affiliation{JILA, National Institute of Standards and Technology and the University of Colorado, Boulder, Colorado 80309, USA}
\affiliation{Department of Physics, University of Colorado, Boulder, Colorado 80309, USA}

\author{K. W. Lehnert}
\email{konrad.lehnert@jila.colorado.edu}
\affiliation{JILA, National Institute of Standards and Technology and the University of Colorado, Boulder, Colorado 80309, USA}
\affiliation{Department of Physics, University of Colorado, Boulder, Colorado 80309, USA}
\date{\today}

\begin{abstract}
Quantum fluctuations constitute the primary noise barrier limiting cavity-based axion dark matter searches. In an experiment designed to mimic a real axion search, we employ a quantum-enhanced sensing technique to detect a synthetic axion-like microwave tone at an unknown frequency weakly coupled to a resonator, demonstrating a factor of 5.6 acceleration relative to a quantum-limited search for the same tone. The acceleration comes from increases to both the visibility bandwidth and the peak visibility of a detector. This speedup is achieved by dynamically coupling the resonator mode to a second (readout) mode with balanced swapping and two-mode squeezing interactions. A small fractional imbalance between the two interaction rates yields further scan rate enhancement and we demonstrate that an 8-fold acceleration can be achieved.

\end{abstract}

\maketitle

\section{Introduction}
\label{sec:intro}

Quantum fluctuations intrinsic to the measurement of a cavity's electromagnetic field \cite{caves1982quantum} are a major source of background noise in searches for weak signals \cite{LIGOsqueezing,bienfait2016reaching, brubaker2017PRL}. As the main source of background in cavity-based axion searches \cite{sikivie1985experimental,bradley2003microwave}, they are the primary limitation inhibiting a comprehensive search of the axion parameter space. The figure of merit for cavity-based axion detectors (haloscopes) is the spectral scan rate. This is defined as the rate at which a search can scan through frequency space while probing with sufficient confidence to resolve or exclude an axion with a given coupling to electromagnetism. It would take thousands of years for existing quantum-limited haloscopes to scan the 1-10 GHz frequency band at benchmark DFSZ coupling \cite{palkenThesis,zhitnitsky1980,dfs1981}.

The scan rate depends on two quantities: the characteristic axion-sensitive bandwidth of the detector and the peak visibility of a potential axion signal. Quantum-enhanced sensing techniques hold promise for increasing these quantities and for facilitating the otherwise prohibitively time- and resource-expensive search. Recently, squeezing has been implemented in an axion search, doubling detector scan rate relative to a search at the quantum limit \cite{backes2021quantum, malnou2019}. In that experiment, one quadrature of the measurement-induced noise was squeezed below the level of vacuum fluctuations, thereby accelerating the search by widening the axion-sensitive bandwidth of the haloscope. Achieving further scan rate enhancement with this technique is challenging due to limitations associated with transporting fragile squeezed states through lossy directional elements in the measurement chain \cite{malnou2019}. 

In this article, we demonstrate a method of scan rate enhancement that circumvents the main limitation of the squeezing technique and achieves both further bandwidth increase and peak visibility improvement, enabling a factor of 5.6 acceleration in the detection of a synthetic axion-like microwave tone. Rather than squeezing the measurement noise, this method depends on amplifying a potential axion signal before it is polluted by measurement noise: an interaction which is enabled by carefully balancing entanglement and swapping interactions between a resonator (cavity) mode and an auxilary readout mode \cite{ceasefire}. To characterize the performance of this quantum-enhanced technique, we perform a realistic acquisition and processing protocol to detect a weak synthetic axion-like microwave tone that has a power spectral density (PSD) which is approximately 1\% of the PSD expected from vacuum fluctuations. By operating the detector with and without the quantum enhancement enabled, we demonstrate a factor of 2.36 improvement in signal-to-noise ratio with the same measurement time, corresponding to a 5.6-fold speed-up over the quantum-limited haloscope. Furthermore, we demonstrate additional scan rate enhancement by imbalancing the parametric interactions slightly \cite{lanes2020harnessing, metelmann2022quantum} which enables greater interaction rates and a bandwidth increase. We show that an 8-fold acceleration is achievable with this modification. Further scan rate enhancement, around 20-fold, should be achievable in applications of this technique to a real axion search.

In the next section, we summarize the key features of the quantum-enhanced method. In Sec.\ \ref{sec:visibility}, we describe the prototype device used in this demonstration experiment and we characterize its performance. In Sec.\ \ref{sec:SRE}, we apply the quantum-enhanced method based on balanced interactions to the search for a weak axion-like microwave tone at an unknown frequency and we extract the scan rate enhancement. Finally, in Sec.\ \ref{sec:GCI}, we demonstrate additional scan rate enhancement using the imbalanced method.

\begin{figure*}[]
	\centering
	\includegraphics[width=17.2 cm]{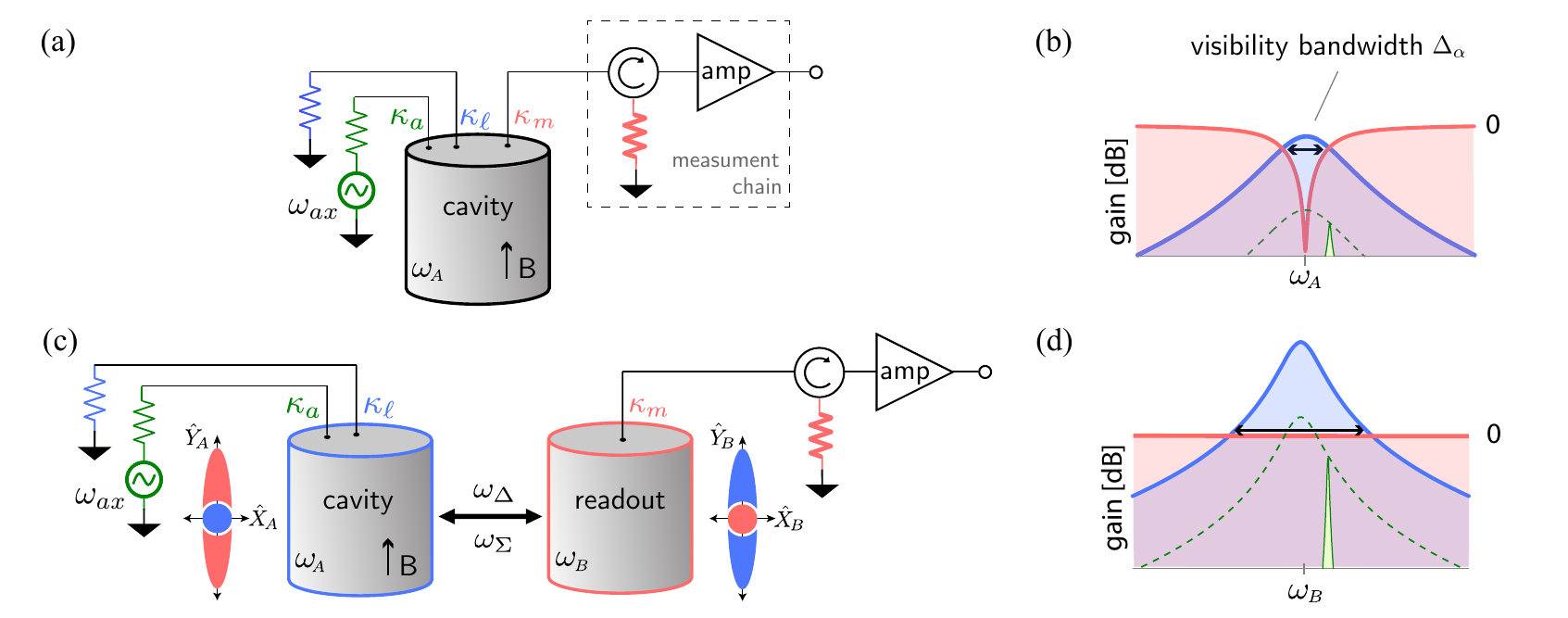} 
	\caption{Comparison between a quantum-limited haloscope and a GC-enhanced haloscope. (a) A microwave network model for a quantum-limited haloscope. The axion signal, modeled by a generator at frequency $\omega_{\mathrm{ax}}$, and the cavity loss, modeled by the blue resistor, are coupled to a cavity centered at $\omega_A$ through fictitious ports at rates $\kappa_a$ and $\kappa_{\ell}$. The outgoing cavity fields couple to the measurement chain at a rate $\kappa_m$ and are routed by a circulator towards a near noiseless amplifier (amp). The circulator shields the cavity from amplifier backaction by dissipating the backaction at a physical termination (bold red resistor). Vacuum noise (measurement noise) sourced from that termination is reflected off the cavity measurement port and couples to the measurement chain, polluting the axion signal. (b) Spectral distribution with respect to vacuum fluctuations of the cavity noise (blue), measurement noise (red), and axion signal (green) at the input of the measurement chain of a quantum-limited haloscope. The green dashed line represents the response to an axion signal at any given frequency, while the narrow green peak represents a potential manifestation of the narrowband axion signal. (c) A microwave network model and phase space diagram for the GC-enhanced haloscope. The measured quadrature, $\hat{Y}_B$, contains amplified axion signal and cavity noise and vacuum level measurement noise, where the size of the fluctuations in the quadrature amplitudes is represented by the blue ellipse and red circle in phase space. (d) At the input of the measurement chain, signal visibility is preserved over a wider bandwidth due to the amplification of the cavity noise and potential axion signal relative to the vacuum-level measurement noise. In contrast to the quantum-limited case, the measurement noise does not vanish on resonance as a result of the balanced parametric interactions.} 
	\label{fig:schematic}
\end{figure*}

\section{Quantum-Enhanced Bandwidth and Visibility Improvement}
\label{sec:ceasefire}
The scan rate of an axion search is limited by vacuum fluctuations arising from both the internal loss of the cavity and the loss external to the cavity. To understand their relative contributions to the scan rate, we consider the effects of these two noise sources as measured at the input of the measurement chain. The internal loss of the cavity (with loss rate $\kappa_{\ell}$) induces cavity noise, which is maximal on cavity resonance. A potential axion signal, which is weakly coupled to the cavity at a rate $\kappa_a$, follows the same path through the haloscope as does the cavity noise, as illustrated in Fig.\ \ref{fig:schematic}(a). Therefore, the ratio of the axion signal PSD to the cavity noise PSD is $\kappa_a/\kappa_\ell$ over the entire spectrum. Measurement noise, arising from loss external to the cavity, dominates off-resonance. The relative spectral contributions of the signal and noise sources are illustrated in Fig.\ \ref{fig:schematic}(b). We define the \emph{visibility} $\alpha$ as the ratio of PSD resulting from a potential axion field to the total noise PSD, and the \emph{visibility bandwidth} $\Delta_\alpha$ as the bandwidth over which the level of cavity noise dominates over the level of measurement noise.

Noiselessly amplifying the cavity noise and potential axion signal together relative to the level of measurement noise increases the visibility bandwidth, as illustrated in Fig.\ \ref{fig:schematic}(d) \cite{ceasefire}. The amplification also results in peak visibility increase, further enhancing the scan rate, as will be discussed in Sec.\ \ref{sec:visibility} and Appx.\ \ref{appendix:visibility_theory}. Ideally, the amplification of the cavity noise and axion signal would occur inside the axion-photon conversion cavity and the amplified fields could be measured directly. However, the magnetic field which enables the axion-photon conversion \cite{sikivie1985experimental} is incompatible with the superconducting circuit elements capable of performing the noiseless amplification. Therefore, the cavity noise and potential axion signal must be transported from the cavity mode (A) to an auxiliary readout mode (B) and amplified there, as depicted in Fig.\ \ref{fig:schematic}(c). 

The two-cavity amplification is achieved by a quantum non-demolition (QND) interaction resulting from two drives. A state swapping interaction at a rate $g_C$, induced by a frequency conversion drive (C) at the difference of the two mode frequencies, $\omega_{\Delta}=|\omega_A-\omega_B|$, is used to continuously exchange the states of the cavity and readout modes. A two-mode squeezing (entanglement) interaction at a rate $g_G$, induced by driving at the sum of the two mode frequencies, $\omega_{\Sigma}=\omega_A+\omega_B$, enables gain (G) and quadrature correlations between the cavity and readout modes \cite{abdo2013nondegenerate,roy2016introduction}. Applying the two drives simultaneously and balancing their interaction rates, $g_C = g_G$, results in a QND interaction given by $\hat{H}\textsubscript{GC} = 2 g_C \hat{X}_{\text{A}} \hat{X}_{\text{B}}$ \cite{ceasefire}, which causes the fields from the cavity mode quadrature $\hat{X}_{\text{A}}$ to be amplified noiselessly at the orthogonal quadrature of the readout mode $\hat{Y}_{\text{B}}$, while $\hat{X}_{\text{A}}$ itself remain unchanged, as shown in the phase space diagram in Fig.\ \ref{fig:schematic}(c). Thus, we can extract the information contained in $\hat{X}_{\text{A}}$ by measuring $\hat{Y}_{\text{B}}$, and the lack of backaction on $\hat{X}_{\text{A}}$ enables a faster measurement rate  compared to that of a quantum-limited haloscope, resulting in bandwidth increase, as shown in Fig.\ \ref{fig:schematic}(d). In the following sections, we refer to this type of quantum-enhanced method as the \emph{GC-enhanced method}. 

\begin{figure*}[]
	\centering
	\includegraphics[width=17.2 cm]{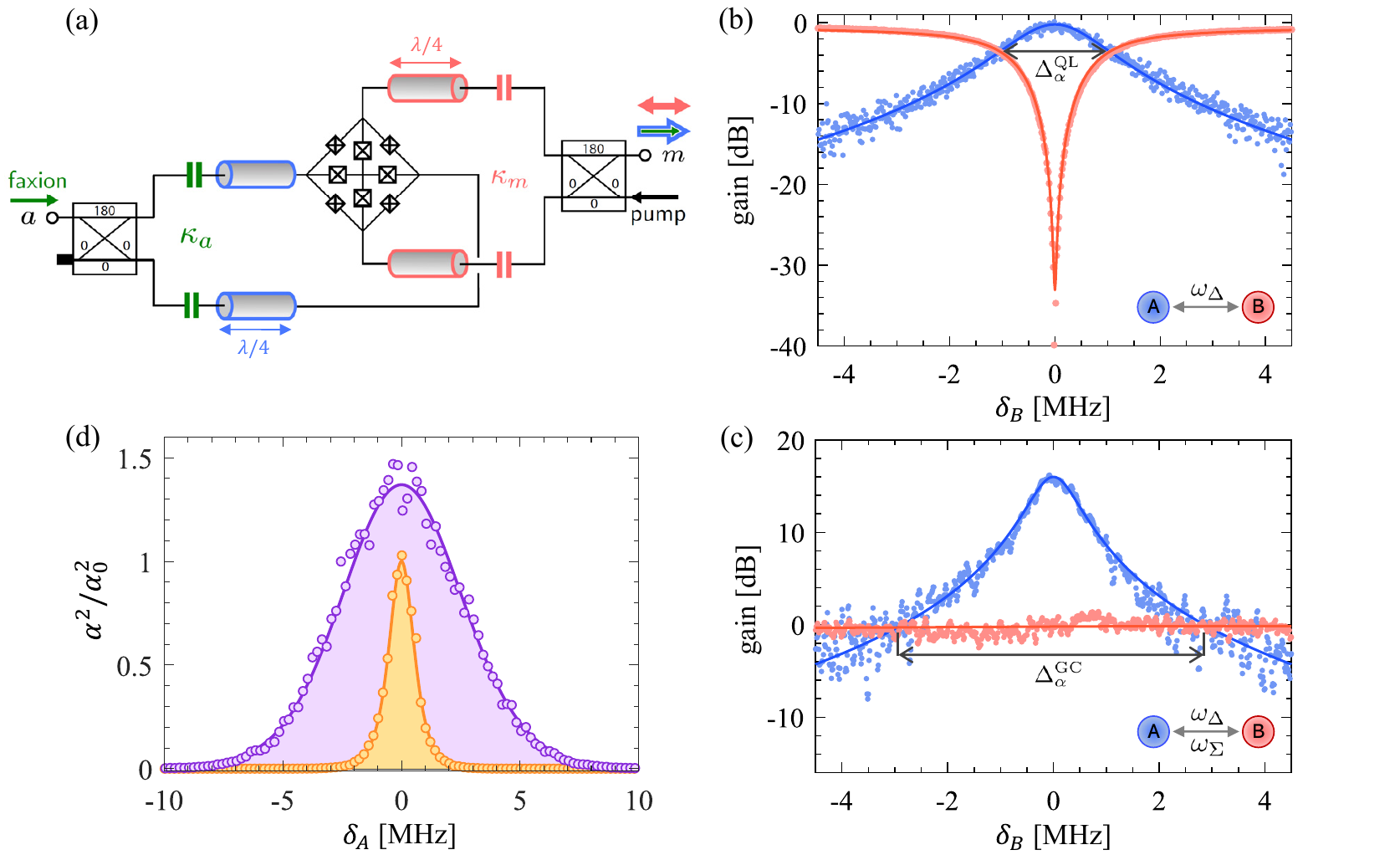} 
	\caption{Experimental demonstration of visibility bandwidth increase. (a) Circuit diagram of a Josephson parametric converter (JPC). The cavity mode (blue) is accessed through the axion port ($a$), and the readout mode (red) is accessed through the measurement port ($m$). (b) Gain of the cavity noise (blue) and measurement noise (red) at the measurement port of the JPC as a function of detuning ($\delta_B$) from $(\omega_\Sigma - \omega_\Delta)/2$ under quantum-limited operation ($g_C = \sqrt{\kappa_m \kappa_\ell}/2$). Fits from the scattering parameters $|S_{m\ell}|^{2}$ and $|S_{mm}|^{2}$ \cite{ceasefire} are included as solid lines, from which we extract the fitting parameters $g_C/2\pi = g_G/2\pi = 7.30\pm0.05$ MHz. (c) Gain from GC-enhanced operation ($g_C = g_G$). Cavity noise is amplified while measurement noise is reflected with unit magnitude, resulting in visibility bandwidth increase ($\Delta_{\alpha}^{\textrm{GC}} > \Delta_{\alpha}^{\textrm{QL}}$). (d) Visibility squared ($\alpha^2$) for the quantum-limited (orange) and GC-enhanced (purple) cases as a function of detuning ($\delta_A$) from $(\omega_\Sigma + \omega_\Delta)/2$. In both cases, $\alpha^2$ is normalized to the peak value $\alpha^2(0)$ of the quantum-limited case. Scan rate enhancement of $5.69\pm0.12$ is extracted from the ratio of the areas enclosed by the purple and orange curves.} 
	\label{fig:gain}
\end{figure*}

\section{Demonstration of bandwidth and visibility improvement}
\label{sec:visibility}
In order to experimentally test the GC-enhanced method, we operate a Josephson parametric converter (JPC) \cite{bergeal2010analog,bergeal2010phase} as a prototype device in a way that mimics a haloscope. The JPC, as illustrated in Fig.\ \ref{fig:gain}(a), consists of two half-wave microstrip resonators coupled by a Josephson ring modulator (JRM), which serves as the three-wave mixing element that enables the swapping and two-mode squeezing interactions. The JPC has three electrical eigenmodes that participate in the JRM: two differential modes which constitute the cavity and readout modes, and a common mode, which is used to apply the $\omega_{\Delta}$ and $\omega_{\Sigma}$ drives that induce the QND interaction \cite{ceasefire}. The cavity mode, which is held at a fixed frequency around $\omega_A/2\pi = 7.454$ GHz in this prototype experiment, has an internal loss rate of $\kappa_{\ell}/2\pi = 960\pm15$ kHz and is weakly coupled to an external port at a rate $\kappa_{a}/2\pi = 1220\pm20$ Hz. This port is accessed via the differential port of a 180$\degree$ hybrid coupler whose two outputs oscillate in anti-phase to excite the cavity mode. It is used to introduce probe tones produced by a microwave generator into the resonator for visibility characterization. The readout mode at $\omega_B/2\pi = 4.98$ GHz is strongly coupled at a rate $\kappa_{m}/2\pi = 20.6\pm1.3$ MHz to the measurement chain through the differential port of another hybrid. The common port of the same hybrid coupler is used to excite the common mode of the JPC and thus to activate the entanglement and swapping interactions.

We operate the JPC under two conditions to demonstrate the visibility bandwidth and peak visibility improvements: one which operates at a quantum-limited scan rate and one which applies the GC enhancement. Here we refer to the quantum-limited scan rate as the maximum achievable scan rate of a critically-coupled haloscope with only vacuum fluctuations entering at its loss and measurement ports and that can measure both quadratures of the field exiting the cavity with the minimum required added noise, or equivalently can measure a single quadrature noiselessly \cite{caves1982quantum}. To achieve this benchmark quantum-limited performance, we apply only the conversion drive (C) to the JPC, thus using it to mimic a single cavity with an effective measurement port coupling rate $\kappa_{m,\mathrm{eff}} = 4g_C^2/\kappa_m$. This approximation is valid in the well-satisfied limit $\kappa_\ell \ll \kappa_m$. For quantum-limited operation, the effective single cavity simulated by the JPC is critically coupled such that $\kappa_{m,\rm{eff}} = \kappa_\ell$. This is achieved by setting $g_C=\sqrt{\kappa_m\kappa_\ell}/2$. To operate in the GC-enhanced mode, both the swap and two-mode squeezing drives are applied at their largest achievable rates before the GC performance becomes unstable as described in Sec.\ \ref{sec:SRE} ($g_C/2\pi=g_G/2\pi = 7.30\pm0.05$ MHz for this device). In both cases, near-noiseless readout is achieved by directing the output of the JPC to a near-noiseless flux-pumped Josephson parametric amplifier (JPA) \cite{yamamoto2008flux,castellanos2008amplification,vijay2011observation,hatridge2011dispersive} which is operated phase sensitively such that in the GC-enhanced case, it amplifies the JPC-amplified quadrature.


We first demonstrate visibility bandwidth increase by measuring the scattering parameters of the JPC in both transmission and reflection. Although there is no accessible port associated with the internal loss of the cavity, the scattering parameter for the cavity noise in transmission, $|S_{m{\ell}}|$, may be inferred from the axion signal scattering parameter $|S_{ma}|$ with  $|S_{m{\ell}}|=\sqrt{\kappa_{\ell}/\kappa_{a}}|S_{ma}|$. We observe unit transmission of the cavity noise on resonance in the quantum-limited case [blue in Fig.\ \ref{fig:gain}(b)], and 16 dB of phase-preserving gain, corresponding with 22 dB of phase-sensitive gain, in the GC-enhanced case [Fig.\ \ref{fig:gain}(c)]. Measuring in reflection off the measurement port, $|S_{mm}|$, we observe near zero reflection on resonance in the quantum-limited case [red in Fig.\ \ref{fig:gain}(b)] as a result of critically coupling ($\kappa_{m,\text{eff}}=\kappa_\ell$), and we observe unit reflection over the entire bandwidth in the GC-enhanced case [Fig.\ \ref{fig:gain}(c)] as a result of the matched interaction rates $g_C = g_G$. The gain experienced by the cavity noise relative to the level of measurement noise in the GC-enhanced case results in a wider visibility bandwidth compared with the quantum-limited bandwidth ($\Delta_{\alpha}^{\textrm{GC}} > \Delta_{\alpha}^{\textrm{QL}}$), as marked by the black arrows.

The other advantage of the GC-enhanced technique, the peak visibility increase, is revealed in the visibility measurement. To perform this measurement, we probe the axion port with a tone generated by a microwave generator, and we discretely step it across the cavity resonance, measuring the visibility $\alpha$ at the measurement port as a function of probe tone detuning from cavity mode resonance $\delta_A$. Plotted in Fig. \ref{fig:gain}(d) is $\alpha^2$ for the quantum-limited and GC-enhanced cases. Given that spectral scan rate scales like $R \propto$ \(\int_{-\infty}^{\infty} \alpha^2 \,\text{d}\delta_A\), a scan rate enhancement of $5.69\pm0.12$ is extracted from the ratio of the areas enclosed by the $\alpha^2$ curves, matching theoretical predictions given the experimentally determined interaction rates extracted from the scattering parameter measurements. 

As demonstrated, scan rate is enhanced by both the visibility bandwidth increase and the peak visibility improvement. The peak visibility improvement arises from the fact that the potential axion signal is amplified before it passes through requisite lossy directional elements in the measurement chain, making it more robust to noise introduced by these lossy components. A more detailed discussion is included in Appx.\ \ref{appendix:visibility_theory}.

\section{Enhanced scan rate in a synthetic axion search}
\label{sec:SRE}

In a real axion search, the goal is to detect a signal that is spectrally broader and several orders of magnitude weaker than the probe tones used to measure visibility in Sec.\ \ref{sec:visibility}. Resolving or excluding such a weak signal at an unknown frequency requires integrating for some time at each cavity tuning step to obtain a power spectrum and then combining the power spectra obtained from many adjacent tuning steps \cite{brubaker2017PRL}. In this section, we use the same measurement setup as described in Sec.\ \ref{sec:visibility} and we demonstrate that the GC-enhanced method still yields significant scan rate enhancement when used to detect a weak synthetic axion-like tone.

To characterize the performance of the GC-enhanced detector in a more realistic search without addressing the added complexity of the magnetic field for axion-photon conversion, we instead inject synthesized fake axion, or ``faxion", tones through the weakly coupled axion port to mimic how an axion signal would couple to a cavity in the presence of a strong magnetic field. The faxion lineshape is achieved by frequency modulating a microwave tone using voltages sampled randomly from a probability distribution which follows the predicted axion spectral distribution \cite{brubaker2017haystac}. We calibrate the power in the faxion tone such that when it is resonant with the cavity mode, it will appear at the measurement port with a peak power spectral density that is roughly 1\% of the level of vacuum fluctuations when measured in the quantum-limited mode of operation. Further details of the faxion lineshape generation are included in Appx.\ \ref{appendix:data_processing}.

\begin{figure*}[t]
	\centering
	\includegraphics[width=17.2 cm]{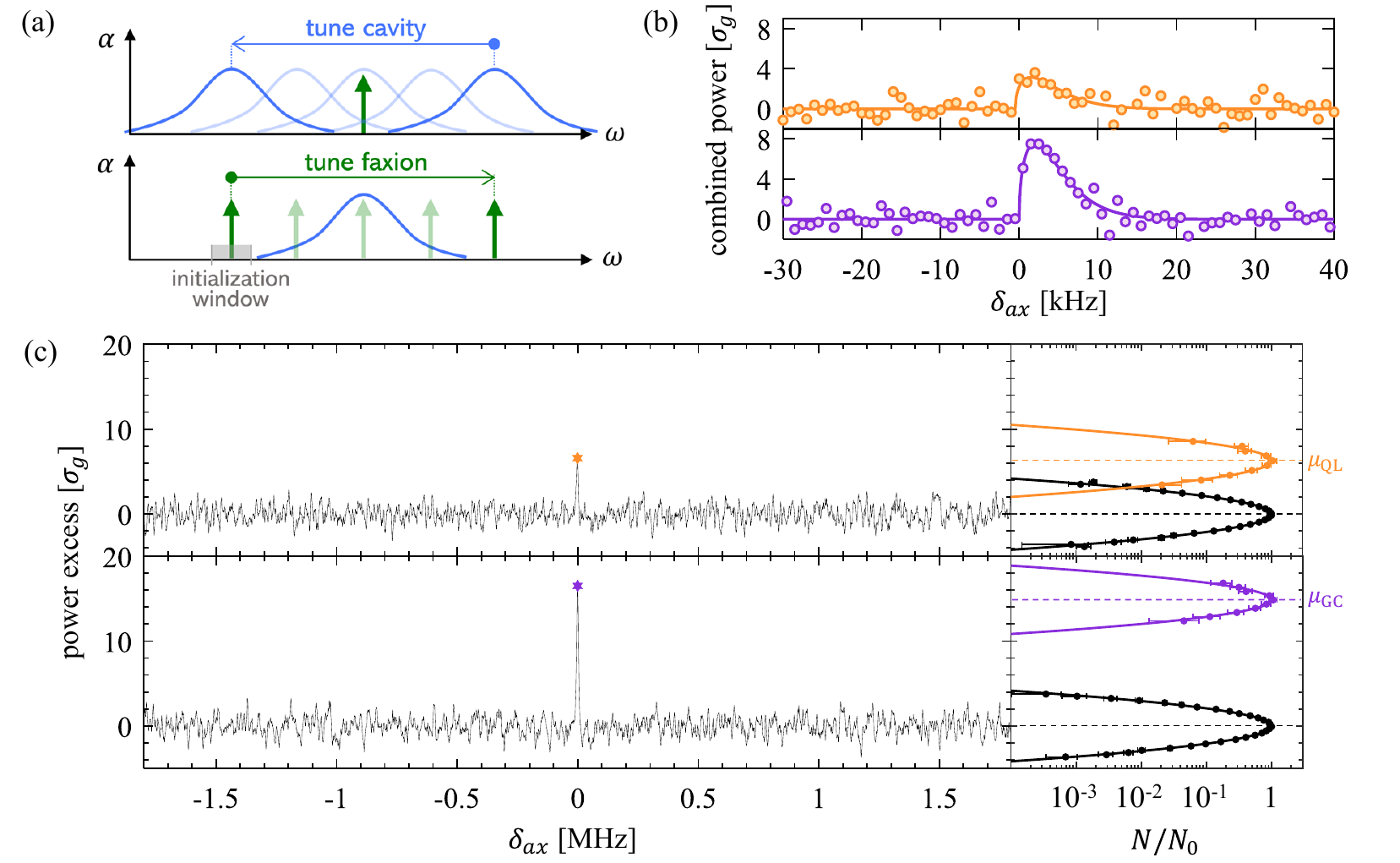} 
	\caption{Scan rate enhancement in detecting a weak signal using the GC-enhanced method. (a) Tuning of the cavity frequency across an axion signal in a real search is simulated by tuning the faxion tone past a fixed-frequency cavity. (b) After combining the power spectra from all tuning steps, the faxion tone can be resolved above the level of noise with a spectral distribution that matches the lineshape of the injected faxion tone, as given by the solid lines. The combined spectrum resulting from the quantum-limited search is plotted in orange while the spectrum from the GC-enhanced search is plotted in purple. (c) Left panels: grand spectra (black) with markers identifying the faxion bins in each case. Right panels: repeating the above process 210 times results in two faxion power excess histograms, normalized to their peak counts $N_0$ and distributed around mean power excesses $\mu_\textrm{QL}$ and $\mu_\textrm{GC}$. A factor of $5.61\pm 0.09$ scan rate enhancement is obtained from $(\mu_\textrm{GC}/\mu_\textrm{QL})^2$. The noise power excess histogram is plotted in black for comparison.} 
	\label{fig:grand}
\end{figure*}

To reduce technical complexity, we simulate the tuning of the cavity across a signal at a fixed but unknown frequency by instead tuning the faxion tone while keeping the cavity mode frequency fixed, as illustrated by Fig.\ \ref{fig:grand}(a). The initial faxion frequency $\omega_{\mathrm{ax}}$ is chosen randomly from within a 1 MHz initialization window outside of the cavity's axion-sensitive band, and the detector is blind to this initial choice. The tone is tuned in 10 kHz steps forward over a 26 MHz window which encompasses the full frequency range over which the cavity is axion-sensitive. At each faxion tuning step, a power spectrum is acquired, resulting in 2601 total spectra. In the data processing, each spectrum is subsequently shifted backward in frequency by the amount the faxion had tuned from the fixed but unknown frequency $\omega_{\mathrm{ax}}$, such that the faxion tones all align at that frequency as if the cavity had been tuned across it \cite{malnou2019}. 

Following the data processing procedure established by prior haloscope experiments \cite{brubaker2017haystac, malnou2019}, we process the shifted spectra to produce a single combined spectrum which can be plotted as a function of detuning $\delta_{ax}$ from the faxion frequency in each of the 2601 spectra. This results in a clear faxion-induced power excess, which follows the expected axion lineshape closely, as plotted in Fig.\ \ref{fig:grand}(b). To further improve the signal-to-noise ratio and resolve the faxion-induced power excess, maximum likelihood estimation is performed on the combined spectrum by accounting for the faxion lineshape. This yields the grand spectrum. Sample grand spectra that result from operating the device in the quantum-limited and GC-enhanced modes are plotted in Fig.\ \ref{fig:grand}(c) with the frequency bins most likely to contain the faxion-induced power excess marked by orange and purple stars.
 
To quantify the scan rate enhancement from the weak signal search, we repeat this faxion injection and detection procedure 210 times with 210 randomly initialized faxion frequencies for both the quantum-limited and GC-enhanced cases. In each case, the faxion bin power excesses from 210 grand spectra are added to a histogram [orange and purple dots in Fig.\ \ref{fig:grand}(d)], resulting in two distributions of faxion-induced power (solid lines). Figure \ref{fig:grand}(d) also displays the noise power distribution (black) obtained from a single grand spectrum which is normally distributed with a mean value of zero and a standard deviation $\sigma_g$. A bin containing faxion-induced power is subject to the same noise fluctuations as the bins containing only noise, but with a faxion-induced power excess which shifts the mean of the distribution away from zero. Therefore, after measurements of equivalent duration, the power in the faxion bins is normally distributed with standard deviation close to $\sigma_g$ and means $\mu_\textrm{QL} = 6.27\pm0.07$ (quantum-limited) and $\mu_\textrm{GC} = 14.85\pm0.06$ (GC-enhanced). The signal-to-noise-ratio (SNR) scales with the square root of the measurement time. Therefore, it would require a $\text{SNR}^2=(\mu_\textrm{GC}/\mu_\textrm{QL})^2 = 5.61$ times longer measurement time for the quantum-limited case to achieve a signal distribution that has the same mean excess as does the GC-enhanced distribution, signifying a factor of $5.61\pm0.09$ scan rate enhancement.

Although the JRM here is not optimized for the Kerr-free operation \cite{sivak2019kerr,chien2020multiparametric,frattini2018optimizing,miano2022frequency}, we note that the scan rate limitation was not, in fact, higher-order parametric processes which would have limited the scan rate enhancement to far below the demonstrated value. The most readily understood higher-order effect is the single-mode squeezing induced by the fourth-order interaction of the JRM, which causes amplified measurement noise and reduced transmission gain, and would have been the primary scan rate enhancement limitation. However, we discovered that this effect can be compensated by detuning the pump frequencies. A more detailed discussion of this compensation procedure will be included in a future publication. 

In this mode of operating the GC amplifier, the limitation to scan rate enhancement comes from the stability of the operating point to drifts in the static flux bias through the JRM. Small fluctuations in the flux bias result in amplified measurement noise, making the performance of the amplifier unstable on time scales much shorter than the duration of the measurement when the device is pumped more strongly. 

\begin{figure}[t]
	\centering
	\includegraphics[width=8.6 cm]{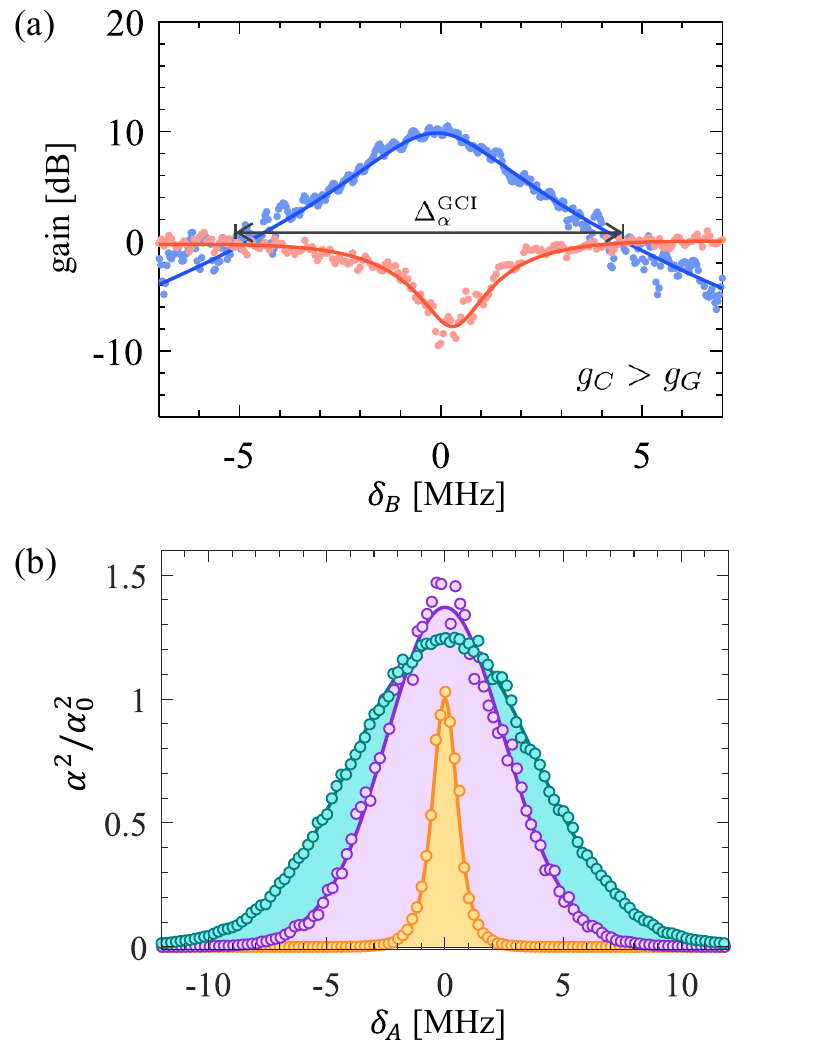}
	\caption{Experimental demonstration of GC-Imbalanced (GCI) operation. (a) Gain of the cavity and measurement noise under the GCI operation. We extract  interaction rates of $g_C/2\pi = 12.25 \pm 0.01\ \mathrm{MHz} > g_G/2\pi = 11.76 \pm 0.01\ \mathrm{MHz}$ from the fits, corresponding with a visibility bandwidth that is larger than the GC bandwidth $\Delta_{\alpha}^{\textrm{GCI}} > \Delta_{\alpha}^{\textrm{GC}}$.
     (b) Visibility squared for the GCI case (teal) relative to the GC and quantum-limited cases. A scan rate enhancement of $8.17 \pm 0.16$ is extracted in the GCI case relative to the quantum-limited case.
     \label{fig:GCI}}
\end{figure}

\section{Further Improvement from Imbalanced Entanglement and Swapping}
\label{sec:GCI}

Although it should be possible to achieve higher balanced GC interaction rates with an active feedback loop to keep the operating point from drifting, here we implement a simpler strategy. Imbalancing the rates slightly such that $g_C \gtrsim g_G$ makes the operating point more stable, allowing for increased interaction rates, a wider visibility bandwidth, and consequently a greater scan rate enhancement. Under GC-imbalanced (GCI) operation \cite{lanes2020harnessing,metelmann2022quantum}, the measurement noise decreases on resonance, rather than being unit reflected over the entire bandwidth, as shown in Fig.\ \ref{fig:GCI}(a). Operating our device at a GC-imbalanced (GCI) point with $g_C/2\pi = 12.25\pm0.01\ \mathrm{MHz} > g_G/2\pi = 11.76\pm0.01$ MHz enables a scan rate enhancement of $8.17 \pm 0.16$, as demonstrated in Fig.\ \ref{fig:GCI}(b). 

Although the interaction rates may be increased further beyond this point by pumping more strongly, additional scan rate enhancement was not achieved. With stronger pumps, we observed that the visibility drops to below the level of the quantum-limited peak visibility, indicating that the amplification induced by the JPC is no longer noiseless. This could be caused by heating from the strong applied pumps or by undesired higher order parametric processes that are not included in our model but contribute more strongly as the pump-induced current across the junctions approaches the critical current. 

To reduce the contribution of these higher order processes, one could tesselate three-wave-mixing dipole elements (SNAILs) on the arms of the JRM \cite{frattini20173}. Alternatively, applying the entanglement interaction between the readout mode and a third ancilla mode (at a different frequency) rather than the cavity mode \cite{li2020broadband} would eradicate some of the higher order parametric processes that may have limited the performance. However, even without optimizing the three-wave mixing element, we predict from theory that maintaining the demonstrated GCI interaction rates but with a reduced cavity mode loss rate equal to that of a typical copper haloscope cavity \cite{backes2021quantum} ($\kappa_\ell/2\pi=100$ kHz) would result in a scan rate enhancement of 20 times.

\section{Conclusion and outlook}
\label{sec:conclusion}
The quantum-enhanced sensing technique demonstrated in this experiment would accelerate an axion search by allowing for fewer total tuning steps or equivalently, for shorter integration time at any given step while achieving the same confidence to resolve or exclude a signal. Next steps towards implementation will involve coupling a microwave cavity immersed in a magnetic field which enables axion-photon conversion to the three-wave-mixing element. Because the superconducting circuitry will need to be spatially separated from the magnetic field, the primary challenge will be achieving sufficient coupling between the cavity and the three-wave mixing element mediated through a transmission line \cite{ceasefire}. With these optimizations, the implementation of this quantum-enhanced technique in a real axion search would mark a significant step towards the feasibility of a comprehensive search of the axion parameter space. 


\section*{Acknowledgements} 
\label{section:acknowledgements}

The authors thank the Devoret group at Yale University for use of their device. The authors additionally thank Daniel Palken and Benjamin Brubaker for helpful discussions. This document was prepared with support from the resources of the Fermi National Accelerator Laboratory (Fermilab), the U.S. Department of Energy, the Office of Science, and the HEP User Facility. Fermilab is managed by Fermi Research Alliance, LLC (FRA), acting under Contract No. DE-AC02-07CH11359. Additionally, this work was supported by Q-SEnSE: Quantum Systems through Entangled Science and Engineering (NSF QLCI Award OMA-2016244) and the NSF Physics Frontier Center at JILA (Grant No. PHY-1734006).

\appendix

\section{Visibility Theory with Measurement Chain Losses}
\label{appendix:visibility_theory}

\begin{figure*}[]
	\centering
	\includegraphics[width=17.2 cm]{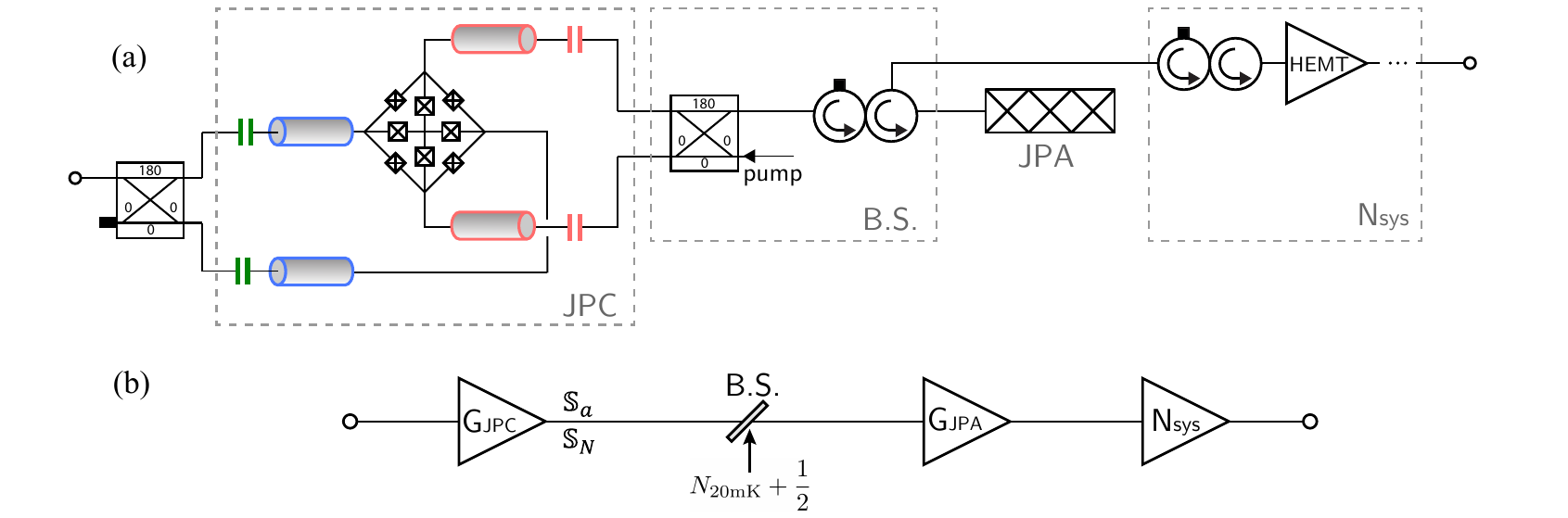} 
	\caption{(a) Simplified microwave network diagram of the experimental setup. The lossy microwave elements between the JPA and the JPC are modeled as a beam splitter (B.S.). (b) The outgoing fields from the JPC $\mathbb{S}\textsubscript{$a$}$, $\mathbb{S}\textsubscript{$N,\mathrm{meas}$}$ are polluted by thermal and vacuum noise which enter at the lossy directional elements modeled by the B.S., resulting in the decrease of peak visibility.} 
	\label{fig:vis_theory}
\end{figure*}

As mentioned in Sec.\ \ref{sec:visibility}, the GC-enhanced technique improves the peak visibility of a haloscope. In this section, we present a theoretical model to explain this effect by accounting for losses in the measurement chain. As displayed in Fig.\ \ref{fig:vis_theory}, we model the loss between the JPC and the JPA as a beam splitter which allows quantum noise comprising thermal and vacuum fluctuations to enter, polluting the signal. In GC operation, the signal is amplified noiselessly by the JPC before it encounters noise entering from this channel. Therefore, it is more robust to this noise than an unamplified signal measured in the quantum-limited case would be.

To model this effect, we write the total noise power spectral density measured at the output of the measurement chain referred to the output of the JPC as
\begin{equation}
\begin{aligned}
 \mathbb{S}_{\mathrm{N,meas}}(\omega) &= \mathbb{S}_{\mathrm{N}}(\omega) 
 + \frac{1-\eta}{\eta}\left(N_{\rm{20mK}}(\omega)+\frac{1}{2}\right)\\
&+\frac{N_{\rm{sys}}}{\eta G_{\rm{JPA}}(\omega)},
\end{aligned}\label{eq:noise}
\end{equation}
where $\mathbb{S}_{\mathrm{N}}$ is the noise power spectral density at the measurement port of the JPC, which comprises the cavity noise and measurement noise, and has an analytical expression given by Eq.\ 14 of reference \cite{ceasefire}. Here, $N_{\rm{20mK}} = (e^{\hbar \omega/k_B \text{T}}-1)^{-1}$ is the Bose occupancy at the fridge base temperature of $\text{T}=20$ mK and the added half quanta in Eq.\ \ref{eq:noise} comes from vacuum fluctuations. The measured phase-sensitive gain of the JPA as a function of frequency is given by $G_{\rm{JPA}}(\omega)$, and $N_{\mathrm{sys}} = 32$ quanta is the total noise added by the system after the JPA. Finally, the energy transmission efficiency of the beam splitter interaction, as modeled in Fig. \ref{fig:vis_theory}, is parameterized by $\eta = 0.9$. As both the JPC and the JPA are operated phase-sensitively, we assume they do not add noise in their measurement of a single quadrature of the signal \cite{caves1982quantum}.

As defined in Sec.\ \ref{sec:ceasefire}, the visibility can be written as the PSD resulting from a potential axion field to the total noise PSD, given by 
\begin{equation}
\begin{aligned}
\alpha(\omega) = \mathbb{S}\textsubscript{$a$}/\mathbb{S}_{\mathrm{N,meas}}(\omega) 
\end{aligned}\label{eq:vis}
\end{equation}
where $\mathbb{S}_{\mathrm{a}}$ is the power spectral density due to an axion signal at the measurement port of the JPC, which has an expression given by reference \cite{ceasefire}. The visibility fits in Fig.\ \ref{fig:gain}(d) and Fig.\ \ref{fig:GCI}(b) are obtained from Eqs.\ \ref{eq:noise} and \ref{eq:vis}.

\section{Synthetic Axion Generation and Data Processing}
\label{appendix:data_processing}
To produce the synthetic faxion tones, we generate a list of voltage samples drawn randomly from a cumulative distribution function (CDF) derived from the axion lineshape \cite{brubaker2017haystac} and we use the list of voltage samples to frequency-modulate a microwave tone. We sample from the CDF to update the faxion frequency at a rate of 1.5 kHz, much slower than the modulation depth which was set to 30 kHz. Meanwhile, the measurement rate at each step (2.4 mHz) was much slower than either of these rates. This means that at any given measurement step, the faxion frequency was updated many times over the course of the integration time, resulting in a signal with a shape that mimics the axion lineshape.

After passing through the full measurement chain, outgoing microwave fields are sent to the radio-frequency (RF) port of a four-port mixer where they are mixed down by a local oscillator (LO) at half the JPA pump frequency. The resulting time domain streams from the in phase (I) and quadrature (Q) channels of the mixer are recorded by an analog to digital converter (ADC, ATS9462) over the course of 0.16 s. This trace is divided into $n=32$ sub-traces with acquisition time $\tau_{\mathrm{aq}}=5$ ms each. Each sub-trace is subsequently Fourier-transformed, yielding a PSD at positive frequencies ($\omega$) which describe detunings from the LO frequency. The PSD at these positive frequencies is subsequently assigned to their negative frequency ($-\omega$) counterparts in the RF. This produces a spectrum that is symmetric about the LO frequency. The 32 transformed and symmetrized sub-traces are then averaged together to produce a single power spectrum, called the ``raw spectrum" which has a frequency-dependent profile. This process is repeated at every faxion tuning step over the 26 MHz range, resulting in 2601 raw spectra. 

In order to identify any power excesses in the data, it is useful to remove the base frequency-dependent profile. To do so, we average the 2601 raw spectra together to get an averaged baseline, and we filter the baseline with a Savitzky-Golay filter to remove all of the small-scale structure from it (features comparable to the axion linewidth and narrower). Then, each raw spectrum is divided by the SG-filtered baseline to remove the frequency-dependent profile while preserving the small-scale structure. Subtracting 1 from these spectra results in 2601 power excess spectra which have mean 0 and standard deviation close to $\sqrt{n \tau_{aq} \Delta_\nu}$, where $\Delta_\nu=200$ Hz is the resolution bandwidth in our measurement. 


The 2601 spectra are then shifted in 10 kHz increments such that the faxion tone in each spectrum lines up with its random initial frequency. This simulates the tuning of a cavity across a fixed-frequency axion tone in a real axion search. Each of the 2601 shifted spectra are then rescaled by the normalized visibility $\alpha(\omega)$ so that frequency bins with higher sensitivity to the faxion are weighted more. The rescaled spectra are then added together to produce a single combined spectrum with mean 0 and standard deviation close to 1, which has a faxion-induced power excess that contains contributions from all of the axion-sensitive frequencies. This procedure results in a clear faxion-induced power excess, which follows the expected axion lineshape closely, as demonstrated in Fig.\ \ref{fig:grand}(b). To further improve the signal-to-noise ratio and determine the frequency bin most likely to contain the faxion, we perform maximum likelihood estimation on the combined spectrum, accounting for the faxion lineshape, which results in the grand spectrum displayed in Fig.\ \ref{fig:grand}(c), with the faxion-induced power excesses marked.

\clearpage

\bibliography{main}

\end{document}